\documentclass[a4paper,preprintnumbers,showpacs,twocolumn,superscriptaddress,nofootinbib,amsmath,amssymb]{revtex4-1}
\usepackage[dvips]{graphics}
\usepackage[hypertex]{hyperref}
\usepackage{color,hyperref}
\usepackage{times}
\usepackage{mathrsfs}
\hypersetup{
  colorlinks=true,        
  linkcolor=blue,         
  citecolor=magenta,      
  filecolor=magenta,      
  urlcolor=blue            
}

\def\beq{\begin{equation}}
\def\eeq{\end{equation}}
\def\bear{\begin{eqnarray}}
\def\ear{\end{eqnarray}}
\def\nn{\nonumber}
\def\L{\mathscr{L}}

\usepackage{enumitem}
\usepackage{graphicx,subfigure}
\usepackage{dcolumn}
\usepackage{bm}
\usepackage{color}

\begin{document}

\title{Electromagnetic perturbations of black holes in general relativity coupled to nonlinear electrodynamics}

\author{Bobir Toshmatov}
\email{bobir.toshmatov@fpf.slu.cz} \affiliation{Institute of
Physics and Research Centre of Theoretical  Physics and
Astrophysics, Faculty of Philosophy \& Science, Silesian
University in Opava, Bezru\v{c}ovo n\'{a}m\v{e}st\'{i} 13,
CZ-74601 Opava, Czech Republic} \affiliation{Ulugh Beg
Astronomical Institute, Astronomicheskaya 33, Tashkent 100052,
Uzbekistan}

\author{Zden\v{e}k Stuchl\'{i}k}
\email{zdenek.stuchlik@fpf.slu.cz} \affiliation{Institute of
Physics and Research Centre of  Theoretical Physics and
Astrophysics, Faculty of Philosophy \& Science, Silesian
University in Opava, Bezru\v{c}ovo n\'{a}m\v{e}st\'{i} 13,
CZ-74601 Opava, Czech Republic}

\author{Jan Schee}
\email{jan.schee@fpf.slu.cz} \affiliation{Institute of Physics and
Research Centre of  Theoretical Physics and Astrophysics, Faculty
of Philosophy \& Science, Silesian University in Opava,
Bezru\v{c}ovo n\'{a}m\v{e}st\'{i} 13,  CZ-74601 Opava, Czech
Republic}

\author{Bobomurat Ahmedov}
\email{ahmedov@astrin.uz} \affiliation{Ulugh Beg Astronomical
Institute,  Astronomicheskaya 33, Tashkent 100052, Uzbekistan}
\affiliation{National University of Uzbekistan, Tashkent 100174,
Uzbekistan}

\begin{abstract}

The electromagnetic (EM) perturbations of the black hole
solutions in general relativity coupled to nonlinear
electrodynamics (NED) are studied for both electrically and
magnetically charged black holes, assuming that the EM
perturbations do not alter the spacetime geometry. It is shown
that the effective potentials of the electrically and magnetically
charged black holes related to test perturbative NED EM fields are
related to the effective metric governing the photon motion,
contrary to the effective potential of the linear electrodynamic
(Maxwell) field that is related to the spacetime metric.
Consequently, corresponding quasinormal (QN) frequencies differ as
well. As a special case, we study new family of the NED black hole
solutions which tend in the weak field limit to the Maxwell field,
giving the Reissner-Nordstr\"{o}m (RN) black hole solution. We
compare the NED Maxwellian black hole QN spectra with the RN black
hole QN spectra.

\end{abstract}

\maketitle

\section{Introduction}\label{sec-intr}

It is well known fact that most of the exact solutions of
Einstein's equations have curvature singularity which is still one
of the unexplained problems of the general relativity. In order to
escape from this inexplicable property of the spacetime, obtaining
the black hole solutions without singularity, i.e., the regular
black hole solutions, has been urged on. One of the simplest ways
to obtain a regular black hole solution is coupling general
relativity to some other fundamental fields such as those
described by the NED. One of the attractive properties of the NED
is the ability to eliminate the curvature singularity from the
black hole
solutions~\cite{ABG:PRL:1998,ABG:PLB:2000,Bronnikov:PRD:2001,Dymnikova:CQG:2004,Fan:PRD:2016,BronnikovPRD:2017}.

It is known that observations~\cite{NarayanAPJ:2006} and analysis
of  data~\cite{LIGOPRL:2016} show that the real astrophysical
black holes are rotating. However, constructing the rotating black
hole solution in general relativity coupled to NED is another
challenge that has not been solved completely. So far, several
authors have made a lot of efforts to take rotating black hole
solutions from existing spherically symmetric ones by using the
Newman-Janis
algorithm~\cite{NewmanJMP:1965,Bambi:PLB:2013,Toshmatov:PRD:2014,Azreg-ainou:PRD:2014,Toshmatov:PRD:2017}
and G\"{u}rses-G\"{u}rsey
algorithm~\cite{GursesJMP:1975,DymnikovaCQG:2015}. Unfortunately,
these rotating solutions are not always representing exact
solutions of the whole set of field equations of the theory.
Namely, the energy-momentum tensor of the rotating regular black
hole solutions obtained by the Newman-Janis algorithm sometimes
does not correspond exactly to the NED
equations~\cite{Rodrigues-Junior:PRD:comment} and such rotating
solutions could be considered only as approximative solutions.
Moreover, induced rotation parameter violates the weak energy
condition of these approximate
solutions~\cite{Bambi:PLB:2013,Toshmatov:PRD:2014,NevesPLB:2014}.

One of the special properties of the NED field is that in such a
field photon does not follow the null geodesics of the background
spacetime metric anymore, instead, it propagates along the null
geodesics of an effective metric which is characterized by the
non-linearities of the
field~\cite{Novello:PRD:2000,Novello:PRD:2001,ObukhovPRD:66,Breto:PRD:2016,deOlivieraCQG:26,StuchlikIJMPD:24}.

In the present paper we focus our attention to the behavior of
the dynamical response of the spherically symmetric, magnetically
and electrically charged black holes representing exact solutions
of coupled Einstein's gravity and NED to small EM perturbations.
Especially, we are going to determine if it is possible to
distinguish the black holes related to the NED from the black
holes related to the standard linear electrodynamics due to their
response to EM perturbations.

Perturbations of black holes imply the study of stability of
their spacetime. The stability of the various black holes in NED
has been studied
in~\cite{BretonPRD:2005,SarbachPRD:67,BretonAP:2015}. Since the
system is open, if the black hole is stable against small
perturbations, it relaxes to its equilibrium state by losing
energy through emitting gravitational, EM or scalar radiation,
depending on the underlying perturbations. The most important part
of this radiation is an intermediate one which is called ringdown
phase that is characterized by a (complex) frequency. Its real and
imaginary parts represent frequency of real oscillations and their
damping rate, respectively. So far, different types of
perturbations of the various regular black holes in NED have been
studied
in~\cite{NomuraPRD:2005,FernandoPRD:2012,FlachiPRD:2013,LiPRD:2013,Toshmatov:PRD:2015,LiEPJC:75,ChaverraPRD:93}.

The paper is organized as follows. In Sec.~\ref{sec-NED} we
present the formalism to construct the electrically and
magnetically charged black hole solutions in GR coupled to the
NED. In Sec.~\ref{sec-new-solution} we present new family of the
magnetically charged black hole solutions. Axial EM perturbations
of the electrically and magnetically charged spherically symmetric
NED black holes and the master equation for them are presented in
Sec.~\ref{sec-perturbation}. QN frequencies, temporal evolution
and stability of EM perturbations of the new obtained NED black
hole solution are studied in Sec.~\ref{sec-qnm-maxwellian}. In
Sec.~\ref{sec-eikonal}, we study and make comparison of the QN
frequencies of the NED black holes and the standard linear
electrodynamics RN black holes in the eikonal (large multipole
number or high frequency) regime. Finally, we present conclusions
implied by our results in Sec.~\ref{sec-conclusion}. In this paper
we mainly use geometrized units $c=1=G$. Furthermore, we adopt
$(-, +, +, +)$ convention for the signature of the metric.

\section{GR coupled to NED}\label{sec-NED}

The action of Einstein's gravity (GR) coupled to the NED is given as
\begin{eqnarray}\label{action}
S=\frac{1}{16\pi}\int d^4x\sqrt{-g}\left(R-\mathscr{L}\right)
\end{eqnarray}
where $g$ is the determinant of the metric tensor $g_{\mu\nu}$,
$R$ is the scalar curvature, and $\L$ represents the Lagrangian
density of the NED field that is function of the electrodynamic
field strength, $\L=\L(F)$, with $F=F_{\mu\nu}F^{\mu\nu}$, where
$F_{\mu\nu}$ is the electrodynamic field tensor that can be
written in terms of a gauge potential as $F_{\mu\nu}=\partial_\mu
A_\nu-\partial_\nu A_\mu$. Definition of the EM field tensor shows
that  $F_{\mu\nu}$ is anti-symmetric and it has only six
independent components.

By neglecting the EM sources, one can write the covariant
equations of motion in the form
\begin{eqnarray}
&&G_{\mu\nu}=T_{\mu\nu},\label{eq-motion1}\\
&&\nabla_\nu\left(\L_F F^{\mu\nu}\right)=0,\label{eq-motion2}
\end{eqnarray}
where the Einstein tensor $G_{\mu\nu}=R_{\mu\nu}-Rg_{\mu\nu}/2$
and $T_{\mu\nu}$ is the energy-momentum tensor of the EM field,
determined by the relation
\begin{eqnarray}\label{em-tensor}
T_{\mu\nu}=2\left(\L_FF_\mu^\alpha F_{\nu\alpha}- \frac{1}{4}g_{\mu\nu}\L\right),
\end{eqnarray}
with $\L_F=\partial_F\L$.

The line element of the static spherically symmetric black hole reads
\begin{eqnarray}\label{line element}
ds^2=-f(r)dt^2+\frac{dr^2}{f(r)}+r^2\left(d\theta^2+\sin^2\theta d\phi^2\right),
\end{eqnarray}
where the lapse function $f(r)$ depends on the NED field. This
line element satisfies the symmetry $G_t^t=G_r^r$. The ansatz for
the EM field can be written in general form as
\begin{eqnarray}\label{ansatz}
\bar{A}_\mu=\varphi(r)\delta_\mu^t-Q_m\cos\theta\delta_\mu^\phi\ ,
\end{eqnarray}
where $\varphi(r)$ is electric potential, while $Q_m$ is the
magnetic charge. Below we construct electrically and magnetically
charged black hole solutions in general relativity coupled to the
NED by the method of Bronnikov~\cite{Bronnikov:PRD:2001}.

\subsection{Electrically charged black hole solution}

The ansatz of the electrically charged black hole solution is
given as $\bar{A}_t=\varphi(r)$. Then, the EM field tensor has
only nonzero component $F_{tr}=-F_{rt}=-\varphi'(r)$. By the
relation $F=F_{\mu\nu}F^{\mu\nu}$ we obtain the EM field strength
as $F=-2\varphi'^2$. Let us consider the metric function $f(r)$
given in the form \bear\label{metric-function}
f(r)=1-\frac{2m(r)}{r}. \ear Then from the Einstein
equation~(\ref{eq-motion1}), we obtain only two independent
equations \bear
&&r^2(\L+4\L_F\varphi'^2)-4m'=0,\label{einstein-el1}\\
&&\L r-2m''=0,\label{einstein-el2}
\ear
By solving above given equations we obtain
\bear
&&\L=\frac{2m''}{r},\label{Lagrangian-el1}\\
&&\L_F=\frac{2m'-rm''}{2r^2\varphi'^2},\label{Lagrangian-el2} \ear
One can see from Eqs.~(\ref{Lagrangian-el1})
and~(\ref{Lagrangian-el2})  that if the mass function does not
depend on radius, $m(r)=M$, the Lagrangian density of the
electrodynamic field vanishes, and we arrive at the solution of
the general relativity itself, i.e. the Schwarzschild metric. From
equations of motion~(\ref{eq-motion2}), the total electric charge
inside the sphere with radius $r$ reads
\bear\label{total-charge}
Q_e=r^2\L_F\varphi'\ .
\ear
By substituting~(\ref{Lagrangian-el2}) to~(\ref{total-charge})
and solving the differential equation, one obtains the electric
potential $\varphi(r)$ in the form
\bear\label{potential-el}
\varphi=\frac{3m-rm'}{2Q_e}+C\ ,
\ear
where $C$ is an integration
constant. If we take the linear electrodynamic field, i.e., the
Maxwell field, our solution reduces to the RN black hole spacetime
with $m(r)=M+Q_e^2/2r$. Then, the electric
potential~(\ref{potential-el}) takes the form $\varphi=Q_e/r$.

By choosing the mass function related to the electric field one
can construct singular and
regular~\footnote{In~\cite{Fan:PRD:2016} it has been shown that in
order for the solution to represent the regular black hole
spacetime, one must choose the mass function so that it satisfies
conditions: $\lim_{r\rightarrow0} m/r^3=finite$,
$\lim_{r\rightarrow0} m'/r^2=finite$, $\lim_{r\rightarrow0}
m''/r=finite$.} black hole solutions in NED.

\subsection{Magnetically charged black hole solution}

The ansatz of the magnetically charged spherically symmetric
black hole spacetime is given by $\bar{A}_\phi=-Q_m\cos\theta$.
Nonzero components of the EM field tensor are
$F_{\theta\phi}=Q_m\sin\theta=-F_{\phi\theta}$. EM field strength
is $F=2Q_m^2/r^4$. Solving the Einstein
equations~(\ref{eq-motion1}) for this case we obtain two
independent equations \bear
&&\L r^2-4m'=0,\label{einstein-mg1}\\
&&4\L_Fq^2-\L r^4+2r^3m''=0,\label{einstein-mg2}
\ear
By solving Eqs.~(\ref{einstein-mg1}) and~(\ref{einstein-mg2}) we obtain
\bear
&&\L=\frac{4m'}{r^2},\label{Lagrangian-mg1}\\
&&\L_F=\frac{r^2(2m'-rm'')}{2Q_m^2},\label{Lagrangian-mg2} \ear If
we assume that the EM field is linear, i.e., the Maxwell  field,
$\L=F$ and $\L_F=1$, then, by solving the above equations we
arrive at the mass function $m=M-Q_m^2/2r$ that represents again
the RN black hole spacetime which is the solution of the
Einstein-Maxwell equations.

\section{New magnetically charged black hole solution}\label{sec-new-solution}

In the paper~\cite{Fan:PRD:2016}, the authors proposed the
formalism presented in section~\ref{sec-NED} and obtained some
black hole solutions in GR coupled to NED. They generalized the
solutions by choosing the Lagrangian density in the form
\bear\label{lagrangian-ned} \L=\frac{4\mu}{\alpha}\frac{(\alpha
F)^{\frac{\nu+3}{4}}}{\left[1+ (\alpha
F)^{\frac{\nu}{4}}\right]^{1+\frac{\mu}{\nu}}}\ , \ear
where $\mu>0$ is a dimensionless constant which characterizes the
strength of nonlinearity of the electrodynamic field, and
$\alpha>0$ is constant parameter which is in the unit of length
squared; $\alpha$ is introduced into theory by the definition
$Q_m=q^2/\sqrt{2\alpha}$. For the magnetically charged nonlinear
electrodynamic field they obtained the solution in the following
form~\cite{Fan:PRD:2016}:
\bear\label{lapse-function}
f(r)=1-\frac{2M}{r}-\frac{2q^3}{\alpha}\frac{r^{\mu-1}}{(r^\nu+q^\nu)^{\frac{\mu}{\nu}}}
\ear
where $q$ is magnetic charge parameter and $\nu>0$ is
dimensionless  constant. $M$ is the pure gravitational mass, let
us say Schwarzschild mass.

However, our calculations show that the Lagrangian
density~(\ref{lagrangian-ned}) gives more general solution in the
form
\bear\label{lapse-function-new}
f(r)=1-\frac{2M}{r}+\frac{2q^3}{\alpha r} -\frac{2q^3}{\alpha}\frac{r^{\mu-1}}{(r^\nu+q^\nu)^{\frac{\mu}{\nu}}}
\ear
Comparing Eqs.~(\ref{lapse-function})
and~(\ref{lapse-function-new}),  one can easily notice that the
difference in the mass functions is the ratio $-q^3/\alpha$ which
cannot be dropped. Mathematically, dropping it also satisfies all
equations, however, dropping of this term is equivalent to $q=0$,
which eliminates the last ratio as well. The asymptotic behaviour
of~(\ref{lapse-function-new}) gives the Arnowitt-Deser-Misner
(ADM) mass of the black hole~\footnote{In the
paper~\cite{Fan:PRD:2016} the ADM mass is given as
$M_{ADM}=M+q^3/\alpha$.} to be $M_{ADM}=M$.

As shown in~\cite{Fan:PRD:2016}, the black hole
solution~(\ref{line element}) with the metric
function~(\ref{lapse-function}) is singular at origin, $r=0$, and
regular only if the pure gravitational mass is neglected, $M=0$,
and $\mu\geq3$. However, the black hole solution with metric
function~(\ref{lapse-function-new}) is singular at $r=0$, even if
$M=0$. The only way to make it regular everywhere in the spacetime
is to assume that the gravitational mass is equal to
\bear\label{reg-condition} M=\frac{q^3}{\alpha}\ , \ear with
$\mu\geq3$. Then one can write the metric
function~(\ref{lapse-function-new}) in the following form:
\bear\label{lapse-function-new1}
f(r)=1-\frac{2Mr^{\mu-1}}{(r^\nu+q^\nu)^{\frac{\mu}{\nu}}}
\ear
Here $M$ is still pure gravitational mass~\footnote{In the
paper~\cite{Fan:PRD:2016} the regular solution also takes the form
of~(\ref{lapse-function-new1}), but $M$ is the electromagnetically
induced mass.}. One may argue that considering the gravitational
mass is constant and playing freely with value of the charge
parameter is impossible, since they are related to each other due
to ~(\ref{reg-condition}). However, fortunately, we have one more
free parameter $\alpha$ which can provide the gravitational mass
to be constant even if the value of charge changes.

\section{Axial EM perturbations of NED black holes}\label{sec-perturbation}

In this section we study axial EM perturbations of black holes in
NED by introducing the axial perturbations into gauge
potential~(\ref{ansatz}) as
\begin{eqnarray}\label{ansatz-perturb}
A_\mu=\bar{A}_\mu+\delta A_\mu\ ,
\end{eqnarray}
considering the perturbations given in the form
\begin{eqnarray}\label{axial-perturb}
\delta A_\mu&=&\sum_{\ell,m}\left(\left[
\begin{array}{c}
0 \\
0\\
\Psi^{\ell m}(t,r)\partial_\phi Y_{\ell m}(\theta,\phi)/\sin\theta\\
-\Psi^{\ell m}(t,r)\sin\theta\partial_\theta Y_{\ell m}(\theta,\phi)\\
\end{array}\right]
\right),
\end{eqnarray}
where $Y_{\ell m}(\theta,\phi)$ is the spherical harmonic function
of degree $\ell$ and order $m$,~\footnote{For the EM perturbations
$\ell=1,2,3,...$ and $m=\pm1,2,\pm3,...,\pm\ell$.} related to the
angular coordinates $\theta$ and $\phi$. Below we study
electrically and magnetically charged black hole cases separately.

\subsection{Electrically charged black hole}

The gauge potential for the electrically charged spherically
symmetric black hole solution is given in the general form
$\bar{A}_\mu=\varphi(r)\delta_\mu^t$.

The nonvanishing covariant components of the EM field tensor of
the 4-potential~(\ref{ansatz-perturb}) with
perturbation~(\ref{axial-perturb}) are given by
\bear\label{covariant-electric}
F_{tr}&&=-\partial_r\varphi,\nn\\
F_{t\theta}&&=\frac{1}{\sin\theta}\partial_t \Psi^{\ell m}\partial_\phi Y_{\ell m}, \nonumber\\
F_{t\phi}&&=-\sin\theta\partial_t \Psi^{\ell m}\partial_\theta Y_{\ell m}, \nonumber\\
F_{r\theta}&&=\frac{1}{\sin\theta}\partial_r \Psi^{\ell m}\partial_\phi Y_{\ell m},\\
F_{r\phi}&&=-\sin\theta\partial_r \Psi^{\ell m}\partial_\theta Y_{\ell m},\nonumber\\
F_{\theta\phi}&&=-\Psi^{\ell
m}\left[\partial_\theta(\sin\theta\partial_\theta Y_{\ell m})
+\frac{1}{\sin\theta}\partial_\phi^2Y_{\ell
m}\right]\nonumber\\&&=-\ell(\ell+1)\Psi^{\ell m}\sin\theta
Y_{\ell m}.\nonumber \ear From the relation
$F^{\mu\nu}=g^{\mu\alpha}g^{\nu\beta}F_{\alpha\beta}$,  we find
the nonvanishing contravariant components of the EM field tensor
\bear\label{contravariant-electric}
&&F^{tr}=\partial_r\varphi,\nn\\
&&F^{t\theta}=-\frac{1}{fr^2\sin\theta}\partial_t \Psi^{\ell m}\partial_\phi Y_{\ell m}, \nonumber\\
&&F^{t\phi}=\frac{1}{fr^2\sin\theta}\partial_t \Psi^{\ell m}\partial_\theta Y_{\ell m}, \nonumber\\
&&F^{r\theta}=\frac{f}{r^2\sin\theta}\partial_r \Psi^{\ell m}\partial_\phi Y_{\ell m},\\
&&F^{r\phi}=-\frac{f}{r^2\sin\theta}\partial_r \Psi^{\ell m}\partial_\theta Y_{\ell m},\nonumber\\
&&F^{\theta\phi}=-\frac{\ell(\ell+1)}{r^4\sin\theta}\Psi^{\ell
m}Y_{\ell m}.\nonumber
\ear
By
combining~(\ref{covariant-electric})
and~(\ref{contravariant-electric})  and taking only first order
perturbations, we find the EM field strength $F$ in the form
\bear\label{em-strength-electric}
F\approx-2\varphi'^2\ .
\ear
Hereafter, prime denotes the partial derivative with respect to
$r$ ($X'=\partial_rX$). One can see from
(\ref{em-strength-electric}) that in the perturbation of gauge
potential, EM field strength, $\bar{F}$, remains unchanged as
\bear\label{LF-electric}
\L_F=\bar{\L}_{\bar{F}}\ .
\ear
By
inserting~(\ref{contravariant-electric}) into~(\ref{eq-motion2})
we get the relation.
\bear\label{eq-motion20}
\partial_t F^{\mu t}&+&\frac{1}{r^2\L_F}\partial_r\left(r^2\L_F
F^{\mu r}\right)+\frac{1}{\sin\theta}\partial_\theta
\left(\sin\theta F^{\mu\theta}\right)\nn\\&&
+\partial_\phi F^{\mu
\phi}=0,
\ear
For $\mu=t$, we arrive at Gauss's law
\bear\label{gauss}
\varphi=\int\frac{Q_e}{r^2\L_F}dr,
\ear
where
$Q_e$ is total charge inside the sphere with radius $r$.  For the
RN black hole case $\L_F=1$, therefore, $\varphi_{RN}=-Q_e/r$
justifies above relation.

For the case of $\mu=r$, equation~(\ref{eq-motion2}) has infinite
solutions. Finally, for $\mu=\theta$ and $\mu=\phi$ we arrive at
the same equation
\bear\label{electric-eq0}
-\frac{\partial^2\Psi}{\partial t^2}+\frac{f}{\L_F}\left(f\L_F
\Psi'\right)' +f\frac{\ell(\ell+1)}{r^2}\Psi=0\ .
\ear
For
simplicity, we choose the function $\Psi$ in the form
\bear\label{function-sep}
\Psi=\frac{1}{\sqrt{\L_F}}\Phi,
\ear
and
introducing the new radial, so-called tortoise coordinate
$dx=dr/f$, we rewrite the equation~(\ref{electric-eq0}) in terms
of the new wave function and arrive at the well-known
Schr\"{o}dinger-like wave equation
\bear\label{weq-electric}
\left[-\frac{\partial^2}{\partial t^2}+\frac{\partial^2}{\partial
x^2}-V_e(r)\right]\Phi_e(r,t)=0\ ,
\ear
where the effective
potential is given by
\bear\label{eff-pot1}
V_e(r)=f\left[\frac{\ell(\ell+1)}{r^2}-\frac{f\L_F'^2-2\L_F\left(f\L_F'\right)'}{4\L_F^2}\right].
\ear
where $\L_F$ is given by the
expression~(\ref{Lagrangian-el2}). As it has already been pointed
out that $F$ and $\L_F$ depend explicitly and implicitly only on
$r$, respectively. Therefore, one can write the first and second
order radial derivatives of $\L_F$ as $\L_F'=\L_{FF}/F'$ and
$\L_F''=(\L_{FFF}- \L_{FF}F'')/F'^2$, respectively. However, when
the black hole solution is constructed by the means shown in
section~\ref{sec-NED}, one will have a problem on expressing the
Lagrangian density $\L$ explicitly as a function of the EM field
strength $F$. Therefore, in this case it is better to keep the
Lagrangian density $\L$ as a function of $r$ as in
(\ref{Lagrangian-el1}) and (\ref{Lagrangian-el2}). Moreover, here
$\L_{FF}=\L_F'/F'$. For the RN black hole $\L_F=1$ or
$m=M-Q_e^2/2r$ ($f=1-2M/r+Q_e^2/r^2$), and for the other black
holes which are not solution of electrodynamics ($F=0$), we obtain
the well-known potential
\bear\label{potential-rn}
V(r)=f\frac{\ell(\ell+1)}{r^2}.
\ear

\subsection{Magnetically charged black hole}

The ansatz of the black hole with magnetic charge reads
$\bar{A}_\mu=-Q_m\cos\theta\delta_\mu^\phi$. Again we add
perturbations~(\ref{axial-perturb}) to the 4-potential
as~(\ref{ansatz-perturb}) and write the nonzero covariant
components of the EM field tensor
\bear\label{covariant-magnetic}
F_{t\theta}&&=\frac{1}{\sin\theta}\partial_t \Psi^{\ell m}\partial_\phi Y_{\ell m}, \nonumber\\
F_{t\phi}&&=-\sin\theta\partial_t \Psi^{\ell m}\partial_\theta Y_{\ell m}, \nonumber\\
F_{r\theta}&&=\frac{1}{\sin\theta}\partial_r \Psi^{\ell m}\partial_\phi Y_{\ell m},\\
F_{r\phi}&&=-\sin\theta\partial_r \Psi^{\ell m}\partial_\theta Y_{\ell m},\nonumber\\
F_{\theta\phi}&&=\sin\theta\left(Q_m-\ell(\ell+1)\Psi^{\ell
m}Y_{\ell m}\right).\nonumber
\ear
By the relation
$F^{\mu\nu}=g^{\mu\alpha}g^{\nu\beta}F_{\alpha\beta}$ the nonzero
contravariant components of the EM field tensor can be written as
\bear\label{contravariant-magnetic}
&&F^{t\theta}=-\frac{1}{fr^2\sin\theta}\partial_t \Psi^{\ell m}\partial_\phi Y_{\ell m}, \nonumber\\
&&F^{t\phi}=\frac{1}{fr^2\sin\theta}\partial_t \Psi^{\ell m}\partial_\theta Y_{\ell m}, \nonumber\\
&&F^{r\theta}=\frac{f}{r^2\sin\theta}\partial_r \Psi^{\ell m}\partial_\phi Y_{\ell m},\\
&&F^{r\phi}=-\frac{f}{r^2\sin\theta}\partial_r \Psi^{\ell m}\partial_\theta Y_{\ell m},\nonumber\\
&&F^{\theta\phi}=\frac{1}{r^4\sin\theta}\left(Q_m-\ell(\ell+1)\Psi^{\ell
m}Y_{\ell m}\right).\nonumber \ear The EM field strength $F$ up to
the first order perturbation terms
\bear\label{em-strength-magnetic}
F\approx\frac{2Q_m^2}{r^4}-\frac{4Q_m\ell(\ell+1)\Psi^{\ell
m}Y_{\ell m}}{r^4}. \ear One can see
from~(\ref{em-strength-magnetic}) that unlike the case of the
electrically charged black hole, axial perturbations change the EM
field strength. In~(\ref{em-strength-magnetic}), the first term
corresponds to the unperturbed EM field strength, $\bar{F}$, while
the second term is the contribution of the perturbation to the
field strength, $\delta F$, i.e., $F=\bar{F}+\delta F$. Because of
the change in the argument $\bar{F}$, the expression of
$\bar{\L_F}$ has been also changed as \bear\label{LF-magnetic}
\L_F=\bar{\L}_{\bar{F}}+\bar{\L}_{\bar{F}\bar{F}}\delta F\ , \ear
where
$\bar{\L}_{\bar{F}\bar{F}}=\partial_{\bar{F}}^2\bar{\L}=\partial_{\bar{F}}\bar{\L_F}$.
Note that $\bar{F}$ and $\bar{\L}_{\bar{F}}$ depend explicitly and
implicitly only on $r$, respectively, while, $\L_F$ is the
function of all coordinates. Now we rewrite the equation of
motion~(\ref{eq-motion2}) in the following form:
\bear\label{eq-motion3}
\partial_t\left(\L_F F^{\mu t}\right)&+&\frac{1}{r^2}\partial_r\left(r^2\L_F F^{\mu r}
\right)+\frac{1}{\sin\theta}\partial_\theta\left(\sin\theta\L_F
F^{\mu\theta}\right)\nn\\&&+\partial_\phi\left(\L_F F^{\mu
\phi}\right)=0, \ear For the cases $\mu=t$ and $\mu=r$, above
equation have infinite solutions.  Therefore, we consider the
cases $\mu=\theta$ and $\mu=\phi$ which imply the following
equation: \bear\label{electric-eq} -\frac{\partial^2\Psi}{\partial
t^2}&+&\frac{f}{\bar{\L_F}}\left(f\bar{\L_F}
\Psi'\right)'\nn\\&&+f\frac{\ell(\ell+1)}{r^2}\left(1+
\frac{4Q_m^2\bar{\L}_{\bar{F}\bar{F}}}{r^4\bar{\L}_{\bar{F}}}\right)\Psi=0\
, \ear By introducing the new function~(\ref{function-sep}), and
the tortoise coordinate, we arrive again to the wave equation
\bear\label{weq}
\left[-\frac{\partial^2}{\partial t^2}+\frac{\partial^2}{\partial x^2}-V_m(r)\right]\Phi_m(r,t)=0\ ,
\ear
where the effective potential is now given by the expression
\bear\label{mag-potential}
&&V_m(r)=\\&&f\left[\frac{\ell(\ell+1)}{r^2}\left(1+
\frac{4Q_m^2\bar{\L}_{\bar{F}\bar{F}}}{r^4\bar{\L}_{\bar{F}}}\right)
-\frac{f\bar{\L}_{\bar{F}}'^2-2\bar{\L}_{\bar{F}}
\left(f\bar{\L}_{\bar{F}}'\right)'}{4\bar{\L}_{\bar{F}}^2}\right].\nn
\ear If we consider linear (Maxwell) electrodynamics,
$\bar{\L}_{\bar{F}}=1$, and we recover again the
potential~(\ref{potential-rn}).

The EM perturbations of the electrically and magnetically charged
black holes in the linear EM fields are governed by the same
potentials given in (\ref{potential-rn}). On the contrary, the EM
perturbations of both electrically~(\ref{eff-pot1}) and
magnetically~(\ref{mag-potential}) charged black holes in general
relativity coupled to the NED are governed by different potentials
and indicate that the electrodynamic nonlinearity must play an
important role in behaviour of perturbations (at least for EM
perturbations).

\section{QNMs of Maxwellian regular black hole}\label{sec-qnm-maxwellian}

As we mentioned already in section~\ref{sec-new-solution}, by
changing the values of the parameters $\nu$ and $\mu$ one can
construct several different singular and regular black hole
solutions. One of the most interesting case of them is the $\nu=1$
case, in which the NED tends to the Maxwell (linear EM) field in
the weak field regime as \bear \L=4\mu F+ O(F^{5/4}). \ear
Therefore, hereafter, we name $\nu=1$ model as Maxwellian black
holes. In this section we study the QNMs of the EM perturbations
of these black holes. To see an effect of the NED, we compare the
results with those related to the RN black holes.

\begin{figure}[h]
\centering
\includegraphics[width=0.43\textwidth]{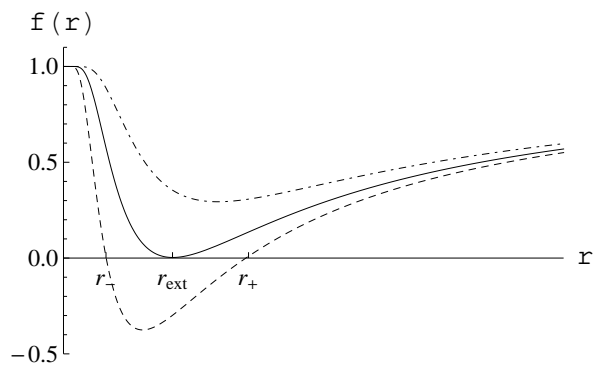}
\caption{\label{fig-lapse} Radial dependence of the metric
function $f(r)$ given by~(\ref{lapse-function}). Here we set
$M=1$. Dashed curve represents black hole, solid curve represents
extreme black hole, and dotdashed curve represents no-horizon
spacetimes.}
\end{figure}

As other regular black holes or RN black hole spacetimes, the
Maxwellian regular black holes also have two horizons: inner
$r_-\geq0$ and outer $r_+\leq2M$, for $q<q_{ext}$, one horizon
$r_-=r_+=r_{ext}$ for $q=q_{ext}$, or no horizon (naked
singularity for the RN spacetime) for $q>q_{ext}$ -- see
Fig.~\ref{fig-lapse}. For the RN black hole: $q_{ext}=M$ and
$r_{ext}=M$. For the Maxwellian regular black hole with $\mu=3$:
$q_{ext}\approx0.2963M$ and $r_{ext}\approx0.5926M$, $\mu=4$:
$q_{ext}\approx0.2109M$ and $r_{ext}\approx0.6328M$, $\mu=5$:
$q_{ext}\approx0.1638M$ and
$r_{ext}\approx0.6554M$.~\footnote{These values can be easily
obtained by solving the equations $f=0$ and $f'=0$,
simultaneously.} Note that we are using the same notation $q$ for
the magnetic (or electric) charge of the RN black hole and
magnetic charge of the Maxwellian regular black holes.

\begin{figure}[h]
\centering
\includegraphics[width=0.43\textwidth]{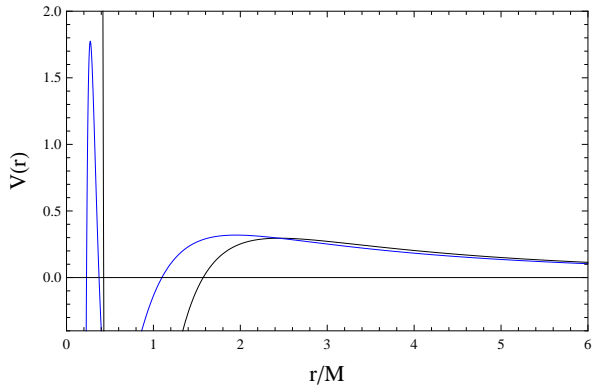}
\caption{\label{fig-veff} Radial dependence of effective
potentials of the EM perturbations of black holes in nonlinear
(Maxwellian black hole with $\mu=3$, blue curve) and linear (RN,
black curve) electrodynamics. Where we set the values of the
charge equal $Q=0.8M$.}
\end{figure}

\begin{figure}[h]
\centering
\includegraphics[width=0.48\textwidth]{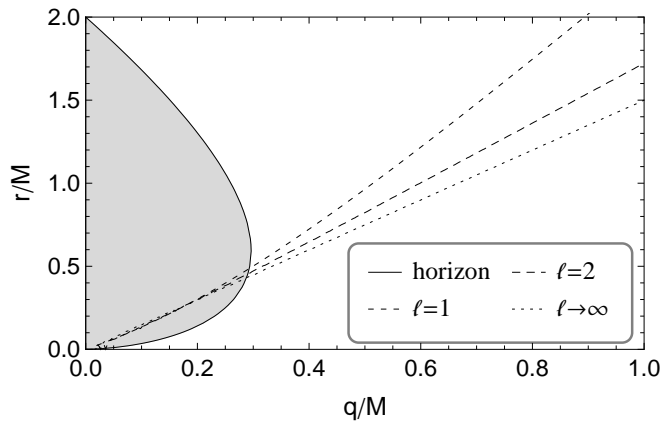}
\caption{\label{fig-rzero} Location of $r_0$ depending on the
charge parameter $q$ for several values of multipole number
$\ell$. The shaded and white regions represent the black hole and
no horizon spacetimes.}
\end{figure}

One can notice that the range of possible values of magnetic
charge parameter of the Maxwellian regular black holes is tighter
than the one of the RN black hole. Therefore, in order to simplify
the comparison, we normalize their values by introducing the new
parameter $Q\equiv q/q_{ext}$. To study the QNMs one should
analyze the effective potentials of the Maxwellian and RN black
holes. Due to the cumbersome length of the effective
potential~(\ref{mag-potential}) for the whole range of parameter
$\mu$ of the Maxwellian regular black
holes~(\ref{lapse-function-new1}), we report here only for the
minimum value of the parameter $\mu$ for the black hole to be
regular, i.e., $\mu=3$ as
\bear\label{potential-maxwellian}
V&&=f\left[\frac{\ell (\ell+1) (2 r-3 q)}{2 r^2 (q+r)}\right.\\
&&\left.\frac{5 q \left(3 r^3 (4 M-3 q)-q r^2 (14 M+3 q)+3
q^4+5q^3 r-4 r^4\right)}{4 r^2 (q+r)^5}\right].\nonumber
\ear
In
Fig.~\ref{fig-veff} we compare the effective potentials of the
RN~(\ref{potential-rn}) and regular Maxwellian black
holes~(\ref{potential-maxwellian}) for the same normalized charge
parameters.
One can see from Fig.~\ref{fig-veff} that outside the event
horizon of the black holes both RN and Maxwellian regular black
holes have very similar potential barriers which tend to zero at
infinity. However, unlike the case in the RN black hole, inside
the event horizon of the Maxwellian black holes, there is another
very narrow potential barrier, located between the inner horizon
$r_-$ and another zero of the effective potential $r_0$, which
depends on $\ell$ and $q/M$. In Fig.~\ref{fig-rzero} dependence of
the location of $r_0$ on $q$ for several values of $\ell$ is
presented.

As one sees from Fig.~\ref{fig-rzero}, the location of $r_0$ is
almost independent of multipole number $\ell$, but it is almost
linearly dependent on the charge $q$. Figs.~\ref{fig-veff}
and~\ref{fig-rzero} show that in the Maxwellian black holes in the
region $r\in[r_-,r_0]\cup[r_+,+\infty)$ there are two potential
barriers. Both of them increase with increasing multipole number
$\ell$. However, their dependence on the charge parameter are
different: with increasing charge parameter, the inner barrier
dramatically decreases, while the outer one increases. Despite
these discussion, one must note that according to the classical
physics, there is nothing coming out from black hole. Therefore,
we do not consider the inner potential barrier. Another really
interesting property of the Maxwellian regular black hole is that
even in the no horizon spacetimes, the effective potential keeps
its barrier form outside $r_0$, i.e., $r\in[r_0,+\infty)$.

\subsection{The temporal evolution of perturbations}

We study the evolution of the EM perturbations by using a
characteristic integration method~\cite{Gundlach:PRD:1994,Rezzolla:CQG} that
involves the light-cone variables: retarded $du\equiv dt-dx$ and
advanced $dv\equiv dt+dx$ time coordinates, with initial data
specified on the two null surfaces $u=u_0$ and $v=v_0$. The wave
equation (\ref{weq}) then takes the form
\begin{equation}
    -4\frac{\partial^2\Phi}{\partial u\partial v}=V(r(u,v))\Phi.
\end{equation}
This equation is solved numerically. The $(u,v)$ space is divided
into finite grid with constant $\Delta$ separating neighboring
points of the grid. The numerical scheme used to solve this
equation reads
\begin{equation}
    \Phi_N=(\Phi_W+\Phi_E)\frac{16-\Delta^2 V_S}{16+\Delta^2 V_S}-\Phi_S
\end{equation}
where the indices $N$, $W$, $E$, and $S$ refer to grid-points
$N\equiv(u,v)$, $W\equiv(u-\Delta,v)$, $E\equiv(u,v-\Delta)$, and
$S\equiv(u-\Delta,v-\Delta)$. In our simulations the initial
perturbation is Gaussian function centered around the point
$x_c$(in tortoise coordinates) and it takes the form
\begin{eqnarray}
    \Phi(t=0,x)&=&A\exp(-(x-x_c)^2/\sigma^2)\nonumber\\
                &=&A\exp(-(v-v_c)^2/\sigma^2)
\end{eqnarray}
since
\begin{equation}
    t=0=\frac{1}{2}(u+v)\,\Rightarrow\, u=-v
\end{equation}
and therefore
\begin{equation}
    x=\frac{1}{2}(v-u)=v
\end{equation}
for $t=0$. At the center of the body, $x=0$, we initially put the
boundary condition $\Phi(u,v)=0$ which is considered along the
line $u=v$ since for
\begin{equation}
    x=0=\frac{1}{2}(v-u)\,\Rightarrow\, u=v.
\end{equation}

\begin{figure}
    \begin{center}
        \includegraphics[scale=0.8]{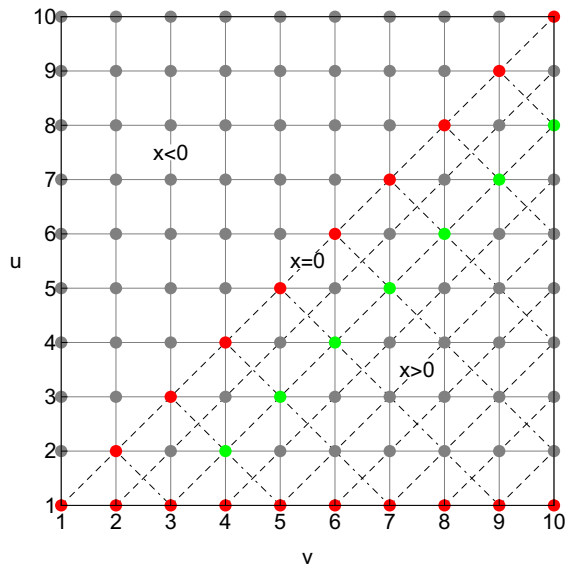}
        \caption{The discretized numerical grid $(u,v)$. The red dots
        represent initial values (horizontal) and boundary values (diagonal).
        The dashed lines correspond to fixed coordinate $x$ while dot-dashed lines
        correspond to fixed coordinate $t$. The green dots here represent the
        solution at a chosen, fixed coordinate $x$. }
    \end{center}
\end{figure}
\begin{figure*}[th]
\centering
\includegraphics[width=0.49\textwidth]{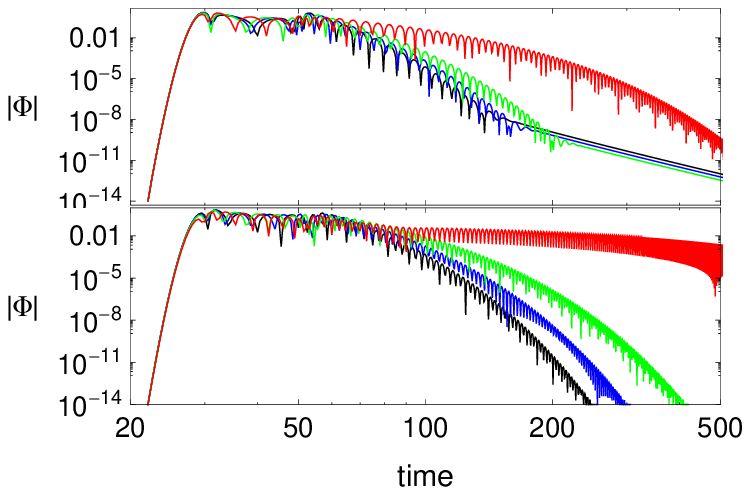}
\includegraphics[width=0.49\textwidth]{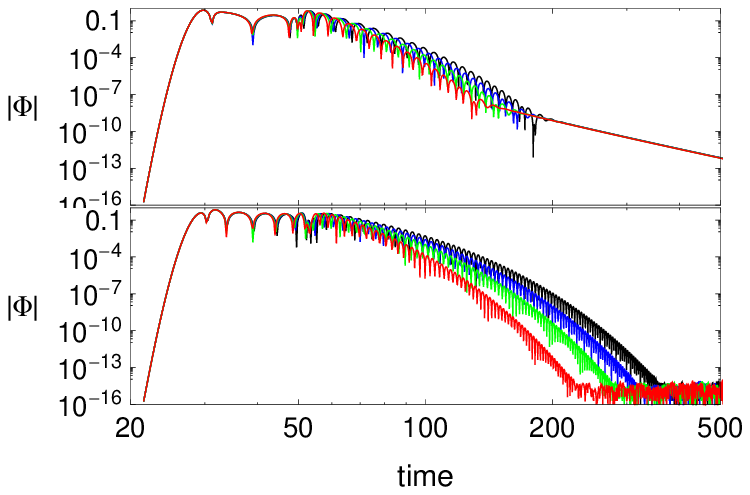}
\caption{Temporal evolution of $\ell=2$ (top panel) and $\ell=4$
(bottom panel) fundamental modes of the EM perturbations of the
Maxwellian regular (left panel) and the RN (right panel) black holes
for the values $Q=0.2$ (black), $Q=0.6$ (blue), $Q=0.8$ (green), and
$Q=0.998$ (red).\label{fig_gw1}}
\end{figure*}

Our tortoise coordinate $x$ is determined from the formula
\begin{equation}
    x=\int_0^r\frac{1}{f(r')}\mathrm{d}r'
\end{equation}
which implies that $x\geq 0$ for $r\geq 0$. We are therefore interested only
in the region where $v\geq u$. In the integration loop  the coordinates $u$ and $v$ are determined by formulas
\begin{equation}
    u=i_u \Delta,\quad\textrm{where}\quad i_u=\{1,2,\dots,N\}
 \end{equation}
 and
\begin{equation}
    v=i_v \Delta,\quad\textrm{where}\quad i_v=\{i_u+1,i_u+2,\dots,N\}.
 \end{equation}
We illustrate the temporal evolution of the EM perturbations of the NED
Maxwellian regular black holes and the RN black holes in Fig.~\ref{fig_gw1}.
One can see from Fig.~\ref{fig_gw1} that the main difference of the
evolution of the EM perturbations in the black holes in the linear and
nonlinear electrodynamics is that an increase in the value of the charge
parameter of the NED Maxwellian regular black holes prolongs perturbations,
while in the linear electrodynamics, it shortens the life of the EM
perturbations. Note that in small and intermediate values of the charge
parameters of the Maxwellian and RN black holes evolution of the EM
perturbations are almost the same.

Moreover, time domain profiles of the EM perturbations of the
Maxwellian regular black holes show that they are stable against EM perturbations.

\subsection{QN frequencies}

In this subsection we calculate QN frequencies by considering
the EM perturbations to be harmonically time dependent as
\bear\label{harmonics}
\Phi(r,t)=\psi(r)e^{-i\omega t}\ ,
\ear
with $\omega=\omega_r+i\omega_i$ being the QN frequency, $\omega_r$
represents frequency of the oscillations, while $\omega_i<0$ ($\omega_i>0$)
represents damping (growing) of these oscillations. Then, the master
equation~(\ref{weq}) takes the new form
\bear\label{weq-new}
\left(\frac{\partial^2}{\partial x^2}+\omega^2-V\right)\psi=0\ .
\ear
Since we are going to study the effect of the nonlinearity of the
electrodynamic field, here $V$ is given by the potentials~(\ref{mag-potential})
and~(\ref{potential-rn}). Since these potentials vanish at the horizon ($x=-\infty$)
and tend to zero at infinity ($x=+\infty$), we choose the boundary condition such
that at horizon (infinity) the wave is purely incoming (outgoing) as
\bear\label{boundary-condition}
\psi\sim e^{\mp i\omega t}, \qquad x\rightarrow\mp\infty.
\ear
\begin{figure}[h]
\centering
\includegraphics[width=0.48\textwidth]{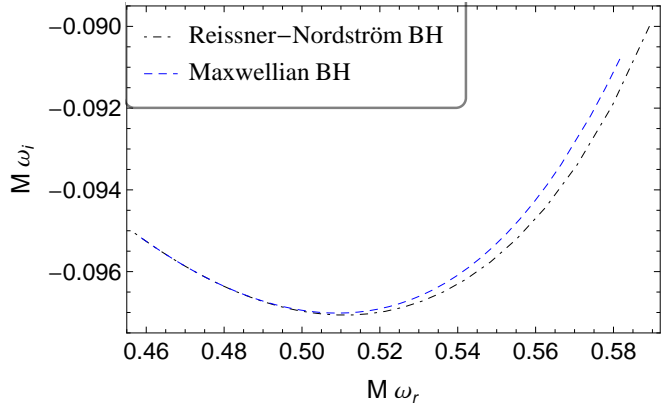}
\caption{\label{fig-qnm} $\ell=2$, $n=0$ QN frequencies of the
EM perturbations of the Maxwellian and RN black holes with the
normalized charge, $Q\in[0,1]$, where $Q=0$ is located at the
junction of the curves which corresponds to the Schwarzschild black hole.}
\end{figure}
\begin{figure}[h]
\centering
\includegraphics[width=0.48\textwidth]{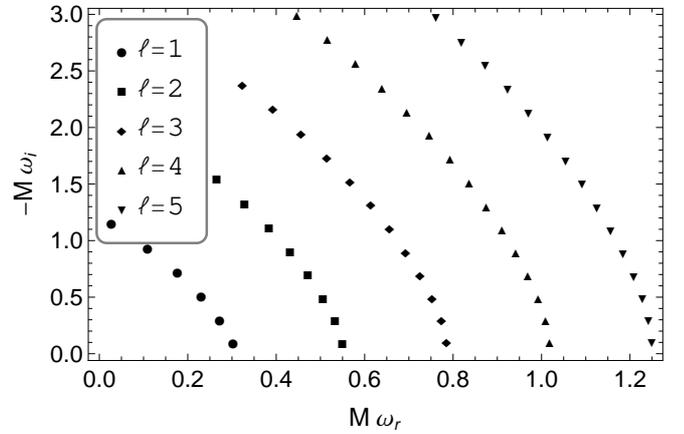}
\caption{\label{fig-nonfundamental} Nonfundamental QN frequencies
of the EM perturbation of Maxwellian regular black hole with $Q=0.8$
for several values of the multipole number $\ell$. The overtone
number, $n$, increases as $n = 0, 1, 2, ...$ from bottom to top.}
\end{figure}
Solving the Eq.~(\ref{weq-new}) with the effective
potential~(\ref{potential-maxwellian}) and boundary
conditions~(\ref{boundary-condition}) analytically is impossible.
Therefore, to solve this equation, we use the well-known
semi-analytical  method, the sixth order WKB
method~\cite{KonoplyaPRD:2003,ToshmatovPRD:2016}. Calculations
show that the QN frequencies of the EM perturbations of the
regular Maxwellian black holes with $\mu=3$ are almost the same as
the ones related to the RN black holes. Therefore, we do not
report all the numerical results; in Fig.~\ref{fig-qnm} in order
to ease comparison, we present some of these results. Moreover, in
Fig.~\ref{fig-nonfundamental} we present some nonfundamental QNMs
of the axial EM perturbation of the regular Maxwellian black
holes.

\section{Eikonal QNMs}\label{sec-eikonal}

In the paper~\cite{CardosoPRD:79} it has been shown that in the
general relativity framework, the QNMs of any stationary,
spherically symmetric and asymptotically flat black holes in any
dimensions are determined in the eikonal (large multipole number)
regime by the parameters of the circular null geodesics, i.e., the
real part of the QN frequencies is determined by angular velocity
of the unstable null geodesics, $\Omega_c$, while the imaginary
part of the QN frequencies is determined by the instability
timescale of the orbit, so called Lyapunov exponent, $\lambda$.
The frequency is thus given by the relation~\cite{CardosoPRD:79}
\bear\label{eikonal1}
\omega=\Omega_c\ell-i\left(n+\frac{1}{2}\right)|\lambda|,
\ear
where $\Omega_c$ and $\lambda$ for the spacetime metric~(\ref{line
element}) are given by the following expressions
\bear
&&\Omega_c=\sqrt{\frac{f_c}{r_c^2}},\label{omega_c}\\
&&\lambda=\sqrt{-\frac{r_c^2}{2f_c}\left(\frac{d^2}{dx^2}
\frac{f}{r^2}\right)|_{r=r_c}},\label{lambda}
\ear
where $x$ is
the tortoise coordinate, $r_c$ is radius of the unstable null
circular orbit which is determined by the solution of equation
$r_cf_c'-2f_c=0$. However, (\ref{eikonal1}) is not universal
feature of all stationary, spherically symmetric and
asymptotically flat black holes in any dimensions, as it has been
shown in~\cite{KonoplyaPLB:2017,KonoplyaJCAP:2017} that these
phenomena are violated in the Einstein-Lovelock theory. Formally,
the same conclusion that the eikonal QNMs of the EM perturbations
are not related to the circular null geodesics, holds for the
metric of the regular black holes considered in this paper.

\begin{figure}[h]
\centering
\includegraphics[width=0.48\textwidth]{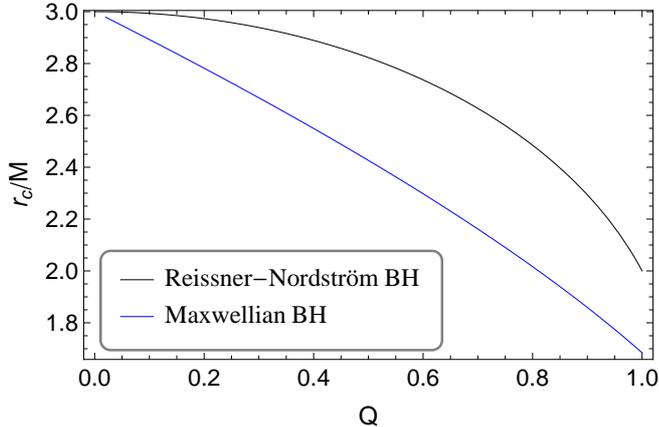}
\caption{\label{fig-photon-orbit} Dependence of radius of circular
unstable null geodesics on the normalized charge parameter of
RN (black) and Maxwellian  regular (blue) black holes.}
\end{figure}

\begin{figure*}[t]
\centering
\includegraphics[width=0.48\textwidth]{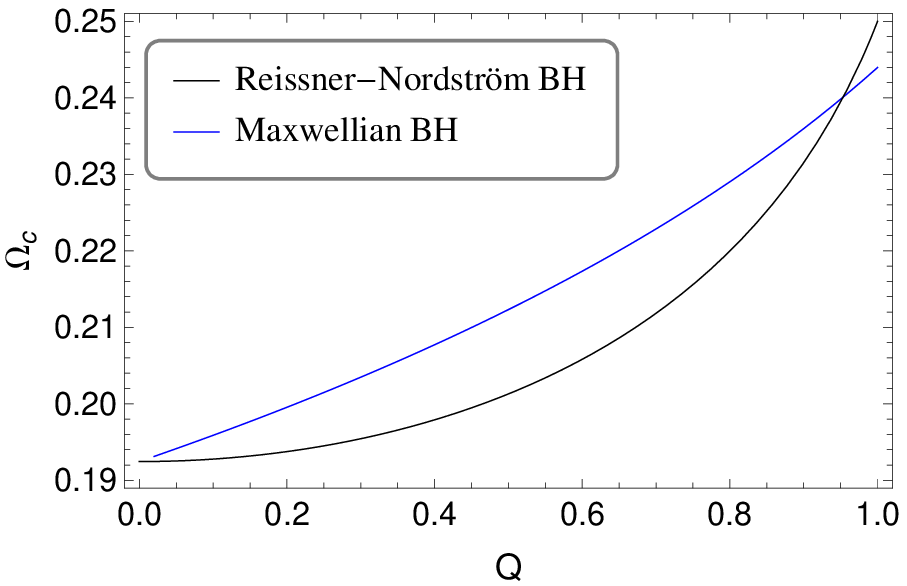}
\includegraphics[width=0.48\textwidth]{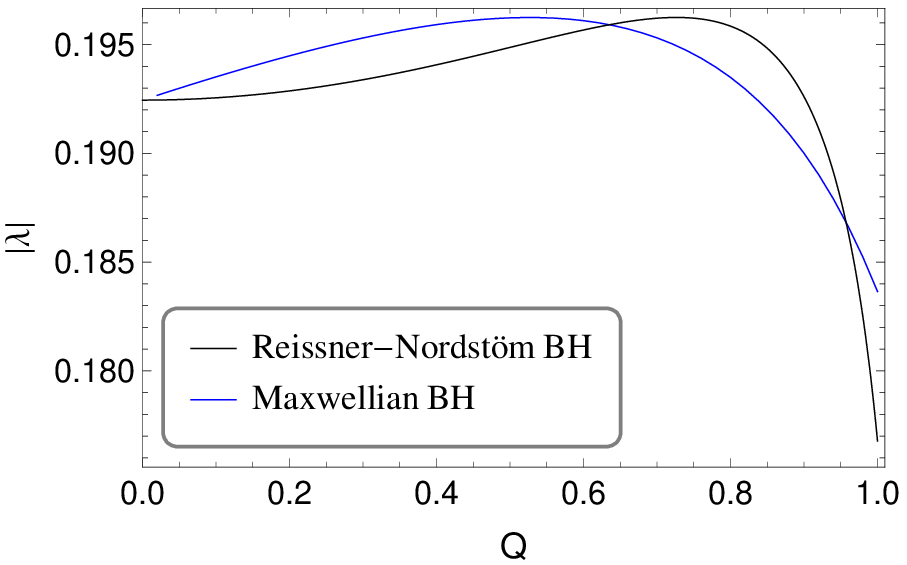}
\caption{\label{fig-eikonal} Dependence of angular velocity and
lyapunov exponent of circular unstable null geodesics on the
normalized charge parameter of RN (black) and Maxwellian  regular
(blue) black holes.}
\end{figure*}

However, in our case the situation is slightly different than in
the Einstein-Lovelock gravity. Let us write the effective
potential of the EM perturbation~(\ref{mag-potential}) for the
large multipole number regime as
\bear\label{potential-eikonal}
V=\ell^2\left[\frac{f}{r^2}\left(1+
\frac{4Q_m^2\bar{\L}_{\bar{F}\bar{F}}}{r^4\bar{\L}_{\bar{F}}}\right)
+O\left(\frac{1}{\ell}\right)\right],
\ear
It is obvious from the
potential~(\ref{potential-eikonal}) that to find the eikonal QNMs,
expressions (\ref{omega_c}) and~(\ref{lambda}) do not work. The
potential~(\ref{potential-eikonal}) corresponds to the one of the
photon motion (NOT NULL GEODESICS) around NED Maxwellian black
holes. It is well know that in the NED, photon does not follow the
null geodesics of original metric, instead, it follows the null
geodesics of the effective optical
metric~\cite{Bronnikov:PRD:2001,Novello:PRD:2000,Novello:PRD:2001,ObukhovPRD:66,Breto:PRD:2016,deOlivieraCQG:26,StuchlikIJMPD:24}.
The effective metric can be constructed as
\bear\label{optical-metric}
g_{eff}^{\mu\nu}=\bar{\L}_{\bar{F}}g^{\mu\nu}-4\bar{\L}_{\bar{F}\bar{F}} \bar{F}_\alpha^\mu\bar{F}^{\alpha\nu}.
\ear
For the magnetically charged Maxwellian black hole with the line
element~(\ref{line element}), covariant components of the
effective metric tensor can be written as the conformal
transformation of the covariant metric tensor
($g_{\mu\nu}^{eff}=\Omega^2g_{\mu\nu}$) as
\bear\label{optical-metric1}
g_{\mu\nu}^{eff}=\left(\bar{\L}_{\bar{F}}+ \frac{4Q_m^2\bar{\L}_{\bar{F}\bar{F}}}{r^4}\right)^{-1}
diag\left\{-g,\frac{1}{h}, r^2, r^2\sin^2\theta \right\}.\nonumber\\
\ear with \bear g=f\left(1+
\frac{4Q_m^2\bar{\L}_{\bar{F}\bar{F}}}{r^4\bar{\L}_{\bar{F}}}\right),
\quad h=\frac{f}{\left(1+
\frac{4Q_m^2\bar{\L}_{\bar{F}\bar{F}}}{r^4\bar{\L}_{\bar{F}}}\right)},
\ear It has been shown that the conformal factor $\Omega^2$ plays
no role in  the EM perturbations~\cite{ToshmatovPRD:96} and null
geodesics~\cite{BambiJCAP:5}. Now one can find the parameters of
circular photon orbit as
\bear
&&\Omega_c=\sqrt{\frac{g_c}{r_c^2}},\label{omega_c1}\\
&&\lambda=\sqrt{\frac{h_c}{2r_c^2}\left(2g_c-r_c^2g_c''\right)}\
,\label{lambda1}
\ear
where $r_c$ is determined by equation
$r_cg_c'-2g_c=0$. In Fig.~\ref{fig-photon-orbit} radii of the
circular unstable null geodesics of the RN and Maxwellian black
holes are depicted.
Numerical calculations show that formula of the eikonal QN
frequencies~(\ref{eikonal1}) works finely in the EM perturbations
of the NED black holes only if instead of the parameters of the
circular null geodesics, the parameters of the circular photon
orbit are used. In Fig.~\ref{fig-eikonal} the angular velocity of
the circular unstable photon orbit~(\ref{omega_c1}) and the
Lyapunov exponent~(\ref{lambda1}) of the RN and Maxwellian regular
black holes are presented.

One can see from Fig.~\ref{fig-eikonal} that in the large
multipole numbers limit, the EM perturbation of the Maxwellian
regular black hole propagates the QNMs with bigger real
frequencies than the RN one with almost the same decaying rates.

\section{Conclusion}\label{sec-conclusion}

In the present paper we demonstrated the formalism to  construct
the electrically and magnetically charged (singular and regular)
black hole solutions in general relativity coupled to the NED. For
our special interest, we constructed the family of new singular
NED black hole solutions which tends to the linear (Maxwell)
electrodynamics in the weak field limit, based on the Lagrangian
density supposed in~\cite{Fan:PRD:2016}. We showed that these
solutions are singular at $r=0$ and unlike the other standard
singular solutions, these solutions are convertible to the regular
ones by the special condition: $M=q^3/\alpha$. As usual regular
black hole spacetimes, these Maxwellian regular black hole
spacetimes also represent black hole, extremal black hole, and no
horizon spacetimes depending on the values of the gravitational
mass and NED parameters.

The main part of this paper is dedicated to the study of the axial
EM perturbations of the general NED black hole solutions
considering the EM perturbations that do not alter the spacetime
geometry. We showed that the EM perturbations of the NED black
holes give different potentials and, consequently, different
results for the QN frequencies, as compared to those related to
the RN black holes in the standard electrovacuum theory. It is
well known that the EM perturbations of the electrically and
magnetically charged black holes in linear electrodynamics (RN)
are isospectral, i.e., they have the same effective potentials and
QN frequencies, however, in the case of the NED black holes,
electrically and magnetically charged black holes have different
potentials and different QNM spectra. As a special case, we
calculated QNMs of the magnetically charged Maxwellian regular
black hole with $\mu=3$ and compared them with the ones of the RN
black holes by normalizing the charge parameter as $Q=q/q_{ext}$
where $Q\in[0,1]$. The analysis of the time domain profile and the
QNM frequencies show that the Maxwellian regular black holes are
stable against EM perturbations.

In the paper~\cite{CardosoPRD:79} it was stated that in the
eikonal (high energy or large multipole number) limit QNMs related
with the unstable circular null geodesics. In this paper we showed
by the EM perturbations of the NED black holes that this claim is
correct in the standard linear electrodynamics, however, it does
not work in the NED, since in the NED photon does not follow the
null geodesics, instead it follows the null geodesics of an
effective metric. We claim that in the eikonal regime, the QNMs of
NED black holes are determined by the unstable circular photon
orbits determined by the effective geometry.

\section*{Acknowledgements}

B.T. would like to thank Roman Konoplya for reading the previous
version of manuscript and providing fruitful discussions and
comments. B.T., Z.S., and J.S. would like to acknowledge the
institutional support of the Faculty of Philosophy and Science of
the Silesian University in Opava, the internal student grant of
the Silesian University (Grant No.~SGS/14/2016) and the Albert
Einstein Centre for Gravitation and Astrophysics under the Czech
Science Foundation (Grant No.~14-37086G). The researches of B.A.
and B.T. are partially supported by Grants No.~VA-FA-F-2-008 and
No. YFA-Ftech-2018-8 of the Uzbekistan Ministry for Innovation
Development, by the Abdus Salam International Centre for
Theoretical Physics through Grant No.~OEA-NT-01. B.T. and B.A.
would like to acknowledge Nazarbayev University, Astana,
Kazakhstan for the warm hospitality through support from ORAU
grant SST~2015021.

\label{lastpage}

\bibliography{Toshmatov_references}

\end{document}